\documentclass[fleqn,usenatbib]{mnras}

\usepackage{newtxtext,newtxmath}

\usepackage[T1]{fontenc}

\DeclareRobustCommand{\VAN}[3]{#2}
\let\VANthebibliography\thebibliography
\def\thebibliography{\DeclareRobustCommand{\VAN}[3]{##3}\VANthebibliography}

\usepackage{graphicx}	
\usepackage{amsmath}	
\usepackage{url}
\usepackage{orcidlink}
\usepackage{widetext}

\title[CSST forecast]{CSST WL preparation I: forecast the impact from non-Gaussian covariances and requirements on systematics-control}

\author[J. Yao et al.]{
	Ji Yao$^{1}$\thanks{E-mail: ji.yao@shao.ac.cn (JY)}\orcidlink{https://orcid.org/0000-0002-7336-2796},
	Huanyuan Shan$^{1,2}$\thanks{E-mail: hyshan@shao.ac.cn (HS)}\orcidlink{https://orcid.org/0000-0001-8534-837X},
	Ran Li$^{3}$,
	Youhua Xu$^{3}$,
	Dongwei Fan$^{3}$,
	Dezi Liu$^{4}$,
	Pengjie Zhang$^{5,6,7}$,
	\newauthor
	Yu Yu$^{5,6}$\orcidlink{https://orcid.org/0000-0002-9359-7170},
	Chengliang Wei$^{8}$,
	Bin Hu$^{9}$,
	Nan Li$^{3}$,
	Zuhui Fan$^{10}$
	Haojie Xu$^{1}$,
	Wuzheng Guo$^{1}$
	\\
	$^{1}$Shanghai Astronomical Observatory (SHAO), Nandan Road 80, Shanghai, China\\
	$^{2}$ University of Chinese Academy of Sciences, Beijing,  China\\
	$^{3}$ National Astronomical Observatory, Chinese Academy of Sciences, Beijing 100101, China\\
	$^{4}$ South-Western Institute for Astronomy Research, Yunnan University, Kunming 650500, Yunnan, PR China\\
	$^{5}$ Department of Astronomy, School of Physics and Astronomy, Shanghai Jiao Tong University, Shanghai, China\\
	$^{6}$ Key Laboratory for Particle Astrophysics and Cosmology
	(MOE)/Shanghai Key Laboratory for Particle Physics and Cosmology,
	China\\
	$^{7}$ Tsung-Dao Lee Institute, Shanghai Jiao Tong University, Shanghai, China\\
	$^{8}$ Purple Mountain Observatory, Chinese Academy of Sciences, Nanjing 210008, China\\
	$^{9}$ Department of Astronomy, Beĳing Normal University, Beĳing 100875, China\\
	$^{10}$ South-Western Institute for Astronomy Research, Yunnan University, Kunming 650500, China\\
}

\date{Accepted XXX. Received YYY; in original form ZZZ}

\pubyear{2015}

\begin{document}
	\label{firstpage}
	\pagerange{\pageref{firstpage}--\pageref{lastpage}}
	\maketitle
	
	\begin{abstract}
		The precise estimation of the statistical errors and accurate removal of the systematical errors are the two major challenges for the stage IV cosmic shear surveys. We explore their impact for the China Space-Station Telescope (CSST) with survey area $\sim17,500\deg^2$ up to redshift $\sim4$. We consider statistical error contributed from Gaussian covariance, connected non-Gaussian covariance and super-sample covariance. {We find the non-Gaussian covariances, which is dominated by the super-sample covariance, can largely reduce the signal-to-noise of the two-point statistics for CSST}, leading to a $\sim1/3$ loss in the figure-of-merit for the matter clustering properties ($\sigma_8-\Omega_m$ plane) and $1/6$ in the dark energy equation-of-state ($w_0-w_a$ plane). We further put requirements of systematics-mitigation on: intrinsic alignment of galaxies, baryonic feedback, shear multiplicative bias, and bias in the redshift distribution, for an unbiased cosmology. The $10^{-2}$ to $10^{-3}$ level requirements emphasize strong needs in related studies, to support future model selections and the associated priors for the nuisance parameters.
	\end{abstract}
	
	
	
	
	\section{Introduction}
	
	Weak gravitational lensing represents a fundamental tool for investigating cosmology, gravity, dark matter, and dark energy \citep{Refregier2003,Mandelbaum2018}. The synergy between weak gravitational lensing (WL) and cosmic microwave background (CMB) observations can yield even more robust results, as it possesses greater constraining power and effectively breaks parameter degeneracies \citep{Planck2018I,DESY3cosmo}. Nonetheless, the tension observed between the cosmic microwave background (CMB) data at redshift $z\sim1100$ and the results obtained from late-time galaxy surveys at $z<\sim1$, possibly caused by unexplained systematic errors or new physics beyond the $\Lambda$CDM cosmological model, poses a significant challenge when employing their combined analysis \citep{Hildebrandt2017,HSC_Hamana2019,HSC_Hikage2019,Asgari2021,Heymans2021,DESY3cosmo,DESY3model,DESY3data,Planck2018I}. A broad range of investigations have been conducted to address the "$S_8$" tension, encompassing diverse systematics \citep{Yamamoto2022,Wright2020b,Yao2020,Yao2017,Kannawadi2019,Pujol2020,Mead2021,DESY3model,Amon2022,Fong2019}, an array of statistical methods \citep{Asgari2021,Joachimi2021,Lin2017b,Harnois-Deraps2021,Shan2018,Sanchez2021,Leauthaud2022,Chang2019,Liu2023}, and the possibility of new physics \citep{Jedamzik2021}. For further reading, we recommend consulting some recent review articles on this topic \citep{Perivolaropoulos2021,Mandelbaum2018}.
	
	To comprehensively address the underlying causes of this tension, various cosmological probes are necessary owing to their distinct sensitivities to systematics and cosmology. Numerous recent observations are preparing to investigate an extended range of redshifts, sky patches, algorithms, and equipment properties. Prominent among these stage IV galaxy surveys are the Dark Energy Spectroscopic Instrument (DESI \citealt{DESI2016a,DESI2016b}), the Legacy Survey of Space and Time (LSST, \citealt{LSST2009}) at the Vera C. Rubin Observatory, Euclid \citep{Euclid2011}, Roman Space Telescope (also known as WFIRST, \citealt{WFIRST2015}) and the China Space Station Telescope (CSST, \citealt{Gong2019}).
	
	In this work, we investigate how accurately CSST can constrain cosmology using its cosmic shear two-point statistics. We test the impact from several important sources of statistical errors and systematical errors. More specifically, we investigate the loss of constraining power in terms of signal-to-noise ratio of the observables and figure-of-merit of the constrained parameters, due to the non-Gaussian covariances \citep{Takada2013,Joachimi2021}. We investigate potential biases in cosmological parameters by analyzing different levels of residual bias in the mitigation of intrinsic alignment \citep{Yao2020,Yao2017,BridleKing,Catelan2001,Hirata2004}, baryonic feedback \citep{Schneider2015,Mead2015,Mead2021,Chen2023}, shear multiplicative bias \citep{Kannawadi2019,Mandelbaum2018HSC,Giblin2021,Liu2021}, and bias in the redshift distribution \citep{Hildebrandt2021,Newman2022,vandenBusch2020,Xu2022,Peng2022}. {These four types of systematics are consided most significant ($\sim10\%$ level contamination in the observable, and $\sim1\sigma$ level potential bias in the cosmology after mitigation, see \citealt{DESY3cosmo,Asgari2021,Li2023}) in the current Stage III surveys, and there impacts could be more significant as the statistical constraining power further improves for Stage IV. We therefore provide calibration requirements for CSST to assist future studies in mitigating those systematics.}
	
	This work is organized as follows. In Section\,\ref{sec theory} we briefly introduce the theoretical predictions to the observables, and the theories for different statistical errors and systematic errors.  In Section\,\ref{sec survey properties} we describe the CSST data we expect. In Section\,\ref{sec results} we forecast the cosmic shear measurements with tomography, and the impact of statistical errors and systematic errors on cosmological parameters. We summarize our findings in Section.\,\ref{sec conclusions}.
	
	
	
	\section{Theory} \label{sec theory}
	
	This section provides a brief review of the cosmic shear two-point statistics theory, how different statistical errors and systematical errors affect the observable, and how we make the forecast with Fisher formalism. We assume $\Omega_k=0$ for spatial curvature, which renders the comoving radial distance and the comoving angular diameter distance identical.
	
	\subsection{Cosmic shear}
	\label{sec cosmic shear}
	
	We employ the lensing convergence auto-correlation in Fourier space, i.e. the lensing angular power spectrum \citep{Asgari2021},\newline
	\begin{equation}
		C^{\kappa\kappa}_{ij}(\ell)=\int_{0}^{\chi_{\rm max}}\frac{q_i(\chi)q_j(\chi)}{\chi^2}  P_{\rm \delta}\left(k=\frac{\ell+1/2}{\chi},z\right)d\chi, \label{eq C^GG}
	\end{equation}
	which is a weighted projection from the 3D {non-linear} matter power spectrum $P_{\rm \delta}(k,z)$ to the 2D galaxy-lensing convergence angular power spectrum $C^{\kappa\kappa}(\ell)$. {In this work, the non-linear matter power spectrum is calculated by halofit \citep{Mead2021,Mead2015,Takahashi2012}.} It also depends on the comoving distance $\chi$, and the lensing efficiency as a function of the lens position (given the distribution of the source galaxies) $q_{\rm s}(\chi)$, which is written as
	\begin{equation}
		q_{\rm s}(\chi_{\rm l}) = \frac{3}{2}\Omega_{\rm m}\frac{H_0^2}{c^2}(1+z_{\rm l})
		\int_{\chi_{\rm l}}^\infty
		n_{\rm s}(\chi_{\rm s})\frac{(\chi_{\rm s}-\chi_{\rm l})\chi_{\rm l}}{\chi_{\rm s}}d\chi_{\rm s}, \label{eq q}
	\end{equation}
	where $\chi_{\rm s}$ and $\chi_{\rm l}$ denote the comoving distance to the source and the lens, respectively, while $n_{\rm s}(\chi_{\rm s})=n_{\rm s}(z)dz/d\chi_{\rm s}$ denotes the distribution of the source galaxies as a function of comoving distance. The ``$\rm s$'' symbols for the source can be replaced by an index for different redshift bins, such as different tomographic bin index $i$ or $j$. {In this work, we consider a flat Universe with $\Omega_k=0$ as weak lensing is not sensitive to the spacial curvature.}
	
	The real-space shear-shear auto-correlation function can be obtained through the Hankel transformation
	\begin{align}
		\xi_{+,ij}(\theta) &= \frac{1}{2\pi}\int_{0}^{\infty}d\ell \ell C^{\kappa\kappa}_{ij}(\ell) J_0(\ell\theta) \label{eq xi+ Hankel}, \\
		\xi_{-,ij}(\theta) &= \frac{1}{2\pi}\int_{0}^{\infty}d\ell \ell C^{\kappa\kappa}_{ij}(\ell) J_4(\ell\theta) \label{eq xi- Hankel},
	\end{align}
	where $J_{0/4}(x)$ is the Bessel function of the first kind with order 0/4. 
	
	Therefore, by observing the correlation $\xi_{+/-,ij}$ or the shear-shear angular power spectrum $C^{GG}_{ij}=C^{\kappa\kappa}_{ij}$, we can derive the constraints on the cosmological parameters through Eq.\,\eqref{eq C^GG}, $P_{\rm \delta}(k)$ and $\chi(z)$. To obtain a precise constraint on cosmology, many sources of statistical errors and systematic errors need to be considered.\newline
	
	\begin{widetext}
		
		\subsection{Covariances} \label{sec cov}
		
		We consider three components of the covariance to account for the statistical error for cosmic shear:
		\begin{equation}
			{\rm Cov_{\rm tot}} = 	{\rm Cov_{Gauss}}+	{\rm Cov_{cNG}}+	{\rm Cov_{SSC}},
		\end{equation}
		namely, the Gaussian covariance, the connected non-Gaussian covariance, and the super-sample covariance.
		
		The Gaussian covariance is based on a common assumption that the fluctuation of the underlying matter field is Gaussian. It is calculated by
		\begin{equation}
			{\rm Cov_{Gauss}}\left(C^{\rm GG}_{ij}(\ell_1),C^{\rm GG}_{mn}(\ell_2)\right)=\frac{\delta_{\ell_1,\ell_2}}{(2\ell+1)\Delta\ell f_{\rm sky}}
			\left[(C^{\rm GG}_{im}+\delta_{i,m}N^{\rm GG}_{ii})(C^{\rm GG}_{jn}+\delta_{j,n}N^{\rm GG}_{jj})
			+(C^{\rm GG}_{in}+\delta_{i,n}N^{\rm GG}_{ii})(C^{\rm GG}_{jm}+\delta_{j,m}N^{\rm GG}_{jj})\right], \label{eq cov G}
		\end{equation}
		
		where $\delta_{\ell_1,\ell_2}$ is the Kronecker delta function; $C^{\rm GG}$ is the shear-shear angular power spectrum; $N^{\rm GG}=4\pi f_{\rm sky}\gamma_{\rm rms}^2/N_{\rm G}$ is the shot noise for $C^{\rm GG}$, where $f_{\rm sky}$ is the fraction of sky of the overlapped area, $N_{\rm G}$ is total number of the galaxies for the source.

		The connected non-Gaussian covariance \citep{Takada2004} provides impact from non-Gaussian distribution of the density field due to late-time non-linear evolution, so that the higher order perturbation enters the covariance. It is calculated by
		\begin{equation}
			{\rm Cov_{cNG}}\left(C^{\rm GG}_{ij}(\ell_1),C^{\rm GG}_{mn}(\ell_2)\right) = \int d\chi \frac{q_i(\chi)q_j(\chi)q_m(\chi)q_n(\chi)}{\chi^6}T_{\rm m}\left(\frac{\ell_1+1/2}{\chi},\frac{\ell_2+1/2}{\chi},a(\chi)\right), \label{eq cov cNG}
		\end{equation}
		where $T_{\rm m}$ is the matter trispectrum, calculated using a halo model formalism \citep{Joachimi2021}. We adopt the NFW halo profile \citep{NFW1996} along with a concentration-mass relation \citep{Duffy2008}, a halo mass function \citep{Tinker2008}, and a halo bias \citep{Tinker2010}.
		
		The super-sample covariance \citep{Takada2013,Takahashi2019,Euclid2023SSC} account for the selection effect of limited observational window, in which the background overdensity can deviate from the ensemble average of the Universe. It is calculated by
		\begin{equation}
			{\rm Cov_{SSC}}\left(C^{\rm GG}_{ij}(\ell_1),C^{\rm GG}_{mn}(\ell_2)\right) = \int d\chi \frac{q_i(\chi)q_j(\chi)q_m(\chi)q_n(\chi)}{\chi^6}
			\frac{\partial P_{\rm \delta}(\ell_1/\chi)}{\partial \delta_{\rm b}}
			\frac{\partial P_{\rm \delta}(\ell_2/\chi)}{\partial \delta_{\rm b}}
			\sigma^2_{\rm b}(\chi), \label{eq cov SSC}
		\end{equation}
		where the derivative of $\partial P_{\rm \delta}/\partial \delta_{\rm b}$ gives the response of the matter power spectrum to a change of the background density contrast $\delta_{\rm b}$, while $\sigma^2_{\rm b}$ denote the variance of the background matter fluctuations in the given footprint. Later we will show the footprint of the CSST cosmic shear observations, which is used to calculate $\sigma^2_{\rm b}$:
		\begin{equation}
			\sigma^2_{\rm b}(\chi) = \frac{1}{A_{\rm eff,\mu}A_{\rm eff,\nu}} \sum_\ell P_{\delta}^{\rm lin}(\frac{\ell}{\chi}) \sum_m a^\mu_{\ell m} a^{\nu *}_{\ell m}.
		\end{equation}
		{The $A_{\rm eff}$ is the effective area for the corresponding spherical harmonic coefficient $a_{\ell m}$ of a probe's mask, with index $\mu$ or $\nu$ representing differet masks, if applicable. And $P_{\delta}^{\rm lin}$ is the linear matter power spectrum.}
		
	\end{widetext}
	
	\subsection{Systematics} \label{sec systematics}
	
	In this work, we consider four major sources of systematics, which can potentially bias the observables at $\sim10\%$ level if unaddressed, in the current Stage III surveys \cite{Asgari2021,DESY3cosmo,HSC_Hikage2019}. Other smaller systematics such as source clustering \citep{Yu2015}, beyond Born approximation \citep{Fabbian2018}, beyond Limber approximation \citep{Fang2020}, and potential CSST-based systematics (similar to \citealt{Euclid2023sys}) are left for future studies.
	
	\subsubsection{intrinsic alignment (IA)}
	
	Weak lensing uses the gravitationally lensed ``optical'' shape $\gamma^{\rm G}$ of the source galaxy to probe the matter and gravity of the lens. The ``dynamical'' shape of a galaxy before being lensed is affected by its local large-scale structures, causing the intrinsic alignment (IA) $\gamma^{\rm I}$, which is therefore a source of systematics. Considering the shape noise $\gamma^{\rm N}$, the overall observed shape reads
	\begin{equation}
		\gamma = \gamma^{\rm G} + \gamma^{\rm I} + \gamma^{\rm N}.
	\end{equation}
	Therefore, in two-point statistics, the target $\left<\gamma^{\rm G}\gamma^{\rm G}\right>$ will be contaminated. In terms of the angular power spectrum,
	\begin{equation}
		C^{\rm \gamma\gamma}_{ij}=C^{\rm GG}_{ij}+C^{\rm IG}_{ij}+C^{\rm GI}_{ij}+C^{\rm II}_{ij}+\delta_{i,j}C^{\rm NN}_{ii}.
	\end{equation}
	Here $C^{\rm \gamma\gamma}$ is the observed angular power spectrum. $C^{\rm GG}$ is the target shear-shear power spectrum, which is identical to the convergence power spectrum as in Eq.\,\eqref{eq C^GG}. $C^{\rm IG}$ and $C^{\rm GI}$ are the shear-IA angular power spectra, which writes
	\begin{equation}
		C^{\rm IG}_{ij}(\ell)=\int_{0}^{\chi_{\rm max}}\frac{n_i(\chi)q_j(\chi)}{\chi^2}  P_{\rm \delta,\gamma^I}\left(k=\frac{\ell+1/2}{\chi},z\right)d\chi, \label{eq C^IG}.
	\end{equation}
	And $C^{\rm II}$ is the IA-IA angular power spectra, given by
	\begin{equation}
		C^{\rm II}_{ij}(\ell)=\int_{0}^{\chi_{\rm max}}\frac{n_i(\chi)n_j(\chi)}{\chi^2}  P_{\rm \gamma^I}\left(k=\frac{\ell+1/2}{\chi},z\right)d\chi, \label{eq C^II}.
	\end{equation}
	
	The 3D matter-IA power spectrum $ P_{\rm \delta,\gamma^I}$ and the 3D IA power spectrum $P_{\rm \gamma^I}$ are based on IA physics. In this work, we use the most widely used NLA model \citep{Asgari2021,DESY3cosmo,HSC_Hikage2019}:
	\begin{align}
		&P_{\rm \delta,\gamma^{\rm I}}=-A_{\rm IA}(L,z)\frac{C_1\rho_{\rm m,0}}{D(z)}P_\delta(k;\chi), \label{eq matter-IA P(k) model} \\
		&P_{\rm \gamma^{\rm I}}=\left(A_{\rm IA}(L,z)\frac{C_1\rho_{\rm m,0}}{D(z)}\right)^2P_\delta(k;\chi), \label{eq IA P(k) model}
	\end{align}
	which are both proportional to the non-linear matter power spectrum $P_\delta$, suggesting that the IA is caused by the gravitational tidal field \citep{Catelan2001,Hirata2004,BridleKing}. It is also affected by $\rho_{\rm m,0}=\rho_{\rm crit}\Omega_{\rm m,0}$ (the mean matter density
	of the universe at $z=0$), the empirical amplitude $C_1=5\times 10^{-14}(h^2M_{\rm sun}/{\rm
		Mpc}^{-3})$ taken from
	\cite{Brown2002}, and $D(z)$, the linear growth
	factor normalized to $D(z=0)=1$. The IA amplitude $A_{\rm IA}$ can be luminosity-dependent \citep{Joachimi2011} or redshift-dependent \citep{Chisari2016,Samuroff2020,Yao2020,Tonegawa2021}.
	
	The IA modeling in Eq.\,\eqref{eq matter-IA P(k) model} and \eqref{eq IA P(k) model} can be replaced by more complicated models such as \cite{Krause2016,Blazek2017,Fortuna2020} for different galaxy samples \citep{Yao2020,Samuroff2020,Zjupa2020}. Its mitigation can alternatively be implemented with extra observables using self-calibration methods \citep{SC2008,Zhang2010,Yao2017,Yao2023KiDS}.
	
	\subsubsection{baryonic feedback}
	
	The modeling of the matter power spectrum $P(k)$ normally uses N-body simulations \citep{Takahashi2012,EuclidEmulator2019}, so that the corresponding density profiles are the dark-matter-only case. The existence of baryonic matter and their associated non-gravitational and powerful process, so-called ``baryonic feedback'', can further change the clustering features in $P(k)$, especially in the small-scales \citep{Jing2006,Schneider2015,Mead2021}. The precise modeling of the baryonic feedback is essential for future cosmological observations \citep{Chen2023,Arico2020,Martinelli2021}.
	
	In this work, we examine the impact of residual baryonic feedback following the baryonic correction model (BCM, \citealt{Schneider2015}):
	\begin{equation}
		P_{\rm BCM}(k,z) = P_{\rm DM}(k,z)F(k,z), \label{eq bcm Pk}
	\end{equation}
	which alters the dark-matter-only power spectrum $P_{\rm DM}$ with a correction term $F$, written as
	\begin{align}
		& F(k,z) = G(k|M_c,\eta_b,z)S(k|k_s), \\
		& G(k|M_c,\eta_b,z) = \frac{B(z)}{1+(k/k_g)^3} + [1-B(z)], \\
		& S(k|k_s) = 1+(k/k_s)^2.
	\end{align}
	Here $G(k|M_c,\eta_b,z)$ represents the suppression from gas dynamics, including AGN feedback, supernovae feedback, etc. $S(k|k_s)$ describes the increase of clustering at small-scale, due to central galaxy stars. Their further expansion for this fitting formula reads:
	\begin{align}
		& B(z) = B_0\left[1+(\frac{z}{z_c})^{2.5}\right]^{-1}, \\
		& k_g(z) = \frac{0.7[1-B(z)]^4\eta_b^{-1.6}}{\rm h/Mpc},
	\end{align}
	with the associated values $z_c=2.3$, $B_0=0.105{\rm log}_{10}\left[\frac{M_c}{\rm Mpc/h}\right]-1.27$, and model parameters $M_c=10^{14.08}$ [Mpc/h] (mass scale), $\eta_b=0.5$ (relation to the escape radius) and $k_s=55$ [h/Mpc] (star component scale).
	
	We consider the un-modeled residual bias in the matter power spectrum in the form of
	\begin{equation}
		\Delta P(k,z) = A_{\rm BCM} \left[ P_{\rm BCM}(k,z) - P_{\rm DM}(k,z) \right],
	\end{equation}
	with an amplitude $A_{\rm BCM}$ to describe how precise we need to understand the true underlying baryonic physics. 
	
	\subsubsection{shear multiplicative bias}
	
	The galaxy shear measurement can suffer from low signal-to-noise (S/N) of the dim galaxies and residuals from the point-spread-function (PSF) deconvolution \citep{Zhang2023}. In the first order, the measured/observed shear can be described as a linear distortion from the true shear:
	\begin{equation}
		\vec{\gamma}^{\rm obs} = (1+\mathbb{M}) \cdot \vec{\gamma}^{\rm true} + \vec{c},
	\end{equation}
	where the 2-component column vector $\vec{\gamma}$ represents the combination of $\gamma_i$ (with $i=1,2$), while $\vec{c}$ represents the corresponding shear additive bias. The $2\times2$ matrix $\mathbb{M}$ contains the shear multiplicative bias $m_{ij}=\partial \gamma^{\rm obs}_i / \partial \gamma^{\rm true}_j$. Generally, for the ``gold'' samples with good shear measurements, we have $|m|\ll 1$ and $|c|\ll |\gamma|$.
	
	In this work, we consider the most common case of a homogeneous and isotropic multiplicative bias $m_{11}=m_{22}=m$, $m_{12}=m_{21}=0$, and negligible additive bias $c_i=0$, similar to the current Stage III observations can achieve \citep{Asgari2021,DESY3data,HSC_Hikage2019}. We note the non-vanishing additive bias can also be removed considering it mainly enters $\xi_+$ (Eq.\,\ref{eq xi+ Hankel}) but not $\xi_-$ (Eq.\,\ref{eq xi- Hankel}), or applying cross-correlations. Similarly, a $z$-dependent $m$ can be further limited with cross-correlations \citep{Liu2021,Yao2023DESI}. In this case, the weak lensing power spectrum is changed by
	\begin{equation}
		\Delta C^{\rm GG} = 2m C^{\rm GG}.
	\end{equation}
	
	\subsubsection{bias in redshift distribution} \label{sec z bias}
	
	As weak lensing requires a large amount of galaxies to suppress the intrinsic shape noise and subtract the cosmological lensing shear signals, photometric redshift (photo-z) is preferred over spectroscopic redshift (spec-z) for its low observational cost. However, the accuracy of photo-z does not satisfy the requirement for the current Stage III and future Stage IV weak lensing surveys, therefore careful calibration of the true redshift distribution is needed \citep{Asgari2021,DESY3cosmo,Wright2020a,Buchs2019,vandenBusch2020,Xu2023,Alarcon2020}.
	
	The main impact of redshift accuracy on cosmology is the mean value of the source redshift distribution. Assume the mean value is biased by $\Delta z$, then the redshift distribution will change from $n(z)$ to $n(z-\Delta z)$, which changes the $n_{\rm s}(\chi_{\rm s}(z_{\rm s}))$ and therefore the lensing efficiency $q_{\rm s}$ through Eq.\,\eqref{eq q} and the theoretical estimation of $C^{\kappa\kappa}$ through Eq.\,\eqref{eq C^GG}.
	
	In this work, we consider a systematic shift in all redshift bins with the same amount $\Delta z$, which can lead to a systematic shift in all cosmological parameters. We note that the bias in redshift distribution is very sample-dependent, therefore the shift is in principle redshift-dependent. However, the z-dependent shift is highly based on the selection function of each specific survey, which is hard to put in the theoretical forecast. Also, the z-dependent bias can be easily identified with cross-correlations \citep{vandenBusch2020,Xu2023} or simply remove a certain z-bin \citep{Asgari2021,Li2023}. So for a concise demonstration, we use an identical z-bias in this work, which also strongly degenerates with the cosmology and is hard to detect.
	
	\subsection{Forecast}
	
	\subsubsection{Fisher formalism} \label{sec Fisher}
	
	We use Fisher matrix \citep{Yao2017,Clerkin2014,Kirk2012,Huterer2006,Coe2009} to pass the statistical uncertainties of CSST observations to the cosmological parameters, to estimate the cosmological constraints and the potential bias from different residual systematics. The Fisher matrix is calculated as:
	\begin{equation}
		\left(F\right) = \left(\frac{\partial \vec{C}^{\kappa\kappa}}{\partial \vec{p}}\right)^T \left({\rm Cov}^{-1}\right) \left(\frac{\partial \vec{C}^{\kappa\kappa}}{\partial \vec{p}}\right), \label{eq Fisher matrix}
	\end{equation}
	where in between $()$ are all matrics. The vector $\vec{C}^{\kappa\kappa}$ is the column data-vector that contains all the $i,j$ combinations and $\ell$ bins in Eq.\,\eqref{eq C^GG}, with total length of $\mathscr{N}_{\rm data}=[(\mathscr{N}_{\rm tomo}+1)\mathscr{N}_{\rm tomo}]/2\times \mathscr{N}_{\ell}$, where $\mathscr{N}_{\rm tomo}$ is the number of tomographic/redshift bins and $\mathscr{N}_{\ell}$ is the number of angular bins. Its partial derivative with respect to the cosmological parameters $\vec{p}$, $\left(\frac{\partial \vec{C}^{\kappa\kappa}}{\partial \vec{p}}\right)$, is therefore a $\mathscr{N}_{\rm data}\times \mathscr{N}_{\rm para}$ matrix, with $\mathscr{N}_{\rm para}$ correspond to the number of cosmological parameters. The inverse of the covariance $ \left({\rm Cov}^{-1}\right)$ uses to the covariance matrices introduced in Sec.\,\ref{sec cov}, and is, therefore, a $\mathscr{N}_{\rm data}\times \mathscr{N}_{\rm data}$ matrix.
	
	By using Eq.\,\eqref{eq Fisher matrix}, we transform the covariance of the data vector to those of the cosmological parameters, with likelihood $-2{\rm ln}\mathscr{L}=\vec{p}^T\left(F\right)\vec{p}$. The $\mathscr{N}_{\rm para}\times \mathscr{N}_{\rm para}$ Fisher matrix $F$ contains direct information of the variance on each parameter $\sigma^2_\alpha=\left(F^{-1}\right)_{\alpha\alpha}$ and the covariance between different parameters $\sigma^2_{\alpha\beta}=\left(F^{-1}\right)_{\alpha\beta}$. The constraining power in the two-parameter-space  can also be evaluated with figure-of-merit (FoM), defined as ${\rm FoM}_{\alpha\beta}=\left[{\rm det}\left(F^{-1}\right)_{\alpha\beta}\right]^{-1/2}$
	
	\subsubsection{Biases due to residual systematics}
	
	To estimate how different residual biases can shift the best-fit cosmology, we first use the introduce systematics in Sec.\,\ref{sec systematics} to estimate a residual bias $\Delta \vec{C}^{\kappa\kappa}$ in the data-vector. Then the corresponding shift in the cosmological parameters $\vec{p}$ can be written as
	\begin{equation}
		\Delta \vec{p} = \left(F^{-1}\right) \left[ \left(\frac{\partial \vec{C}^{\kappa\kappa}}{\partial \vec{p}}\right)^T \left({\rm Cov}^{-1}\right) \left(\Delta\vec{C}^{\kappa\kappa}\right)\right], \label{eq shift}
	\end{equation}
	considering first order approximation to the likelihood \citep{Yao2017,Huterer2006}.
	
	We note that in this work, we aim at getting a general requirement on the systematics, to guide future calibration works on different systematics. For some systematics such as IA and baryonic feedback, as the true model to describe the physics is still unknown, it is trivial to consider the marginalization over their nuisance parameters, due to different model's parameters can have different degeneracies with the cosmological parameters. For any of the systematics mitigation methods, validation with simulation is also a crucial link, which can both estimate the potential residual bias and give priors to the nuisance parameters. Assuming the simulations can give strong enough priors, we no longer need to consider the constraining-power-loss due to marginalization. {We therefore aim at getting the requirements from the residual systematics first, then check how it is affected by considering the marginalization of the nuisance parameters.}
	
	\subsubsection{Signal-to-noise (S/N) definition}
	\label{sec S/N}
	
	The conventional S/N definition uses amplitude fitting \citep{Yao2023DESI}. For a given measurement $\vec{w}_{\rm data}$ and an assumed theoretical model $\vec{w}_{\rm model}$, we fit an amplitude $A$ to the likelihood:
	\begin{equation}
		-2{\rm ln}\mathscr{L}=\left(\vec{w}_{\rm data}-A\vec{w}_{\rm model}\right)^T \left({\rm Cov}^{-1}\right) \left(\vec{w}_{\rm data}-A\vec{w}_{\rm model}\right),
	\end{equation}
	so that a posterior of $A^{+\sigma_A}_{-\sigma_A}$ can be obtained, where $\sigma_A$ is the Gaussian standard deviation for the amplitude. Then the corresponding S/N is $A/\sigma_A$.
	
	Under the frame of Fisher formalism, we have $\vec{w}_{\rm data} = \vec{w}_{\rm model}$ with a single free parameter $A$ for the forecast. Similar to the procedures in Sec.\,\ref{sec Fisher}, one can estimate the S/N is $A/\sigma_A=\sqrt{\left(\vec{w}_{\rm model}\right)^T \left({\rm Cov}^{-1}\right) \left(\vec{w}_{\rm model}\right)}$.
	
	\section{Survey properties} \label{sec survey properties}
	
	\begin{figure}
		\includegraphics[width=\columnwidth]{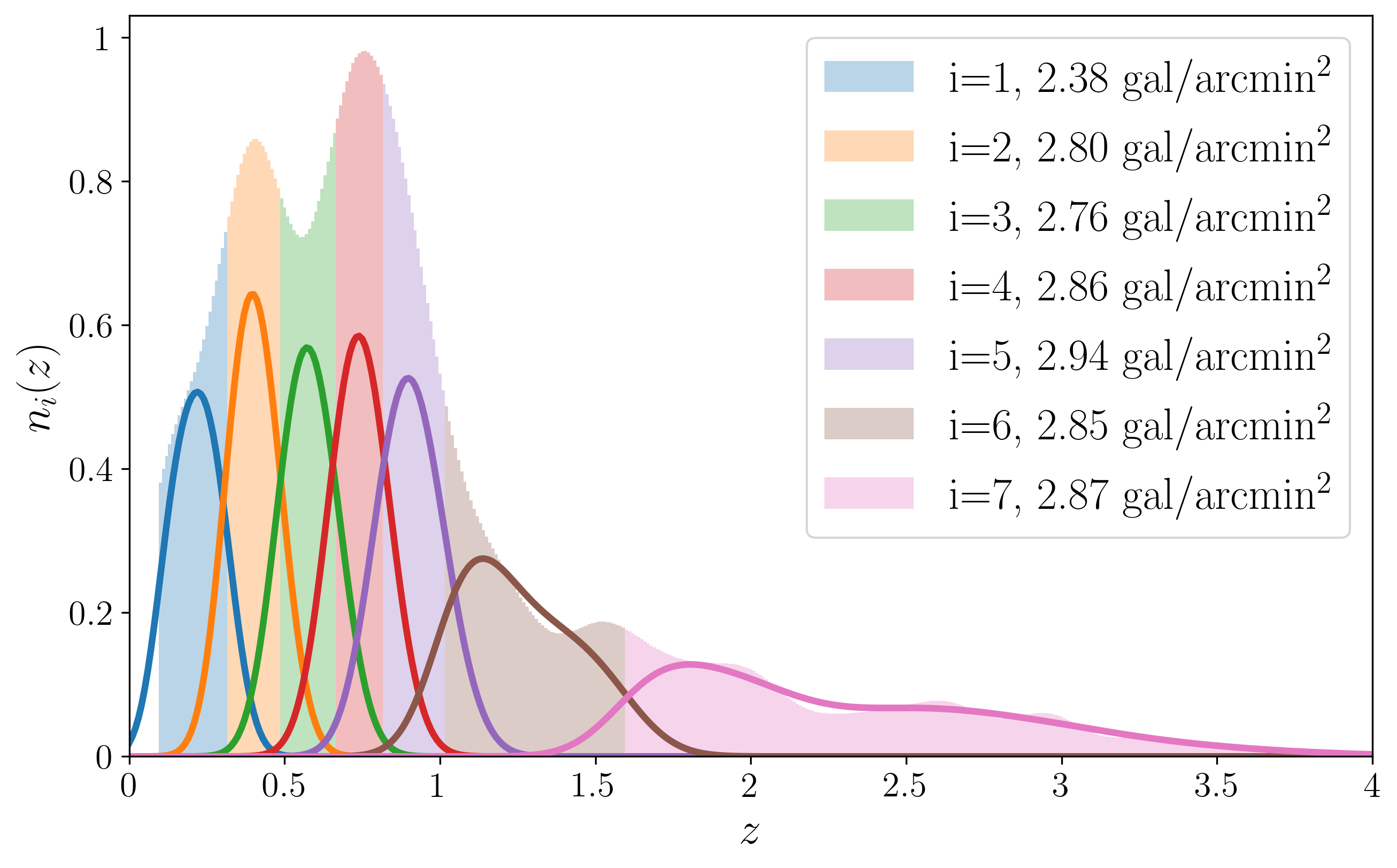}
		\caption{The tomographic redshift distribution for the forecast CSST gold lensing samples. The histogram is the photo-z distribution for different tomographic bins. The solid curves are the associated true-z distributions for each bin for the fiducial analysis, assuming no redshift bias. Galaxies with $z_{\rm p}<0.1$ are removed due to low lensing efficiency.} \label{fig nz}
	\end{figure}
	
	The China Space Station Telescope (CSST) is a space-based project aiming at mapping the Universe with both photometric and slitless spectroscopic observations, covering 17,500 deg$^2$ of the sky. Its imaging survey contains 7 photometric bands (NUV, u, g, r, i, z, y) with wavelength coverage from 255 nm to 1000 nm \citep{Gong2019,Zhan2021}. The $5\sigma$ point sources detection limit in nominal AB magnitude is r $\sim26$ \citep{Cao2018}.
	
	We present the CSST shear catalog properties for this forecast work. For reliable shear and photo-z measurements, we consider a reduction in the galaxy number density from 28 gal/arcmin$^2$ \citep{Gong2019} to 20 gal/arcmin$^2$ \citep{Liu2023} with appropriate S/N cuts. For more detailed studies considering blending and masking \citep{Chang2013}, it will require a PhoSim\citep{Peterson2015}-like simulator that highly mimics the CSST galaxies, which is under development. However, the blending problem is less significant comparing with LSST \citep{Liu2023}.
	
	In Fig.\,\ref{fig nz}, we show the expected redshift distribution of the CSST source galaxies. The photo-z distribution is obtained by applying the CSST observational limits to the COSMOS photo-z galaxies \citep{Cao2018,Ilbert2009}. We divide the galaxies into 7 tomographic bins with (almost) equal numbers of galaxies, for higher total S/N following \cite{Moskowitz2022}. We remove galaxies with photo-z $z_{\rm p}<0.1$ as they contribute little to the total cosmological signals due to low lensing efficiency as in Eq.\,\eqref{eq q}. We assume the true-z follows a Gaussian probability distribution function (PDF) around the photo-z, namely
	\begin{equation}
		p(z|z_{\rm p}) = \frac{1}{\sqrt{2\pi}\sigma_z(1+z)}{\rm exp}\left\{-\frac{(z-z_{\rm p}-\Delta_z)^2}{2[\sigma_z(1+z)]^2}\right\},
	\end{equation}
	where we adapt the photo-z scatter $\sigma_z=0.05$ and photo-z bias $\Delta_z=0$ for the fiducial analysis \citep{Cao2018}. The resulting true-z distributions are also shown as the solid curves in Fig.\,\ref{fig nz}. A non-vanishing photo-z bias $\Delta_z$ is equivalent to a systematic shift in the overall redshift distribution with $\Delta z$, introduced in Sec.\,\ref{sec z bias}.
	
	We consider galaxy shape noise $\gamma_{\rm rms}=0.27$ that is close to the other stage IV surveys \citep{Yao2017}. It will enter the shot noise term $N^{\rm GG}$ as shown in Eq.\,\eqref{eq cov G}. For a more detailed estimation of how much additional statistical error can be introduced from the shear measurement, a more realistic imaging simulation is needed to highly mimic the CSST galaxy properties.
	
	\begin{figure}
		\includegraphics[width=\columnwidth]{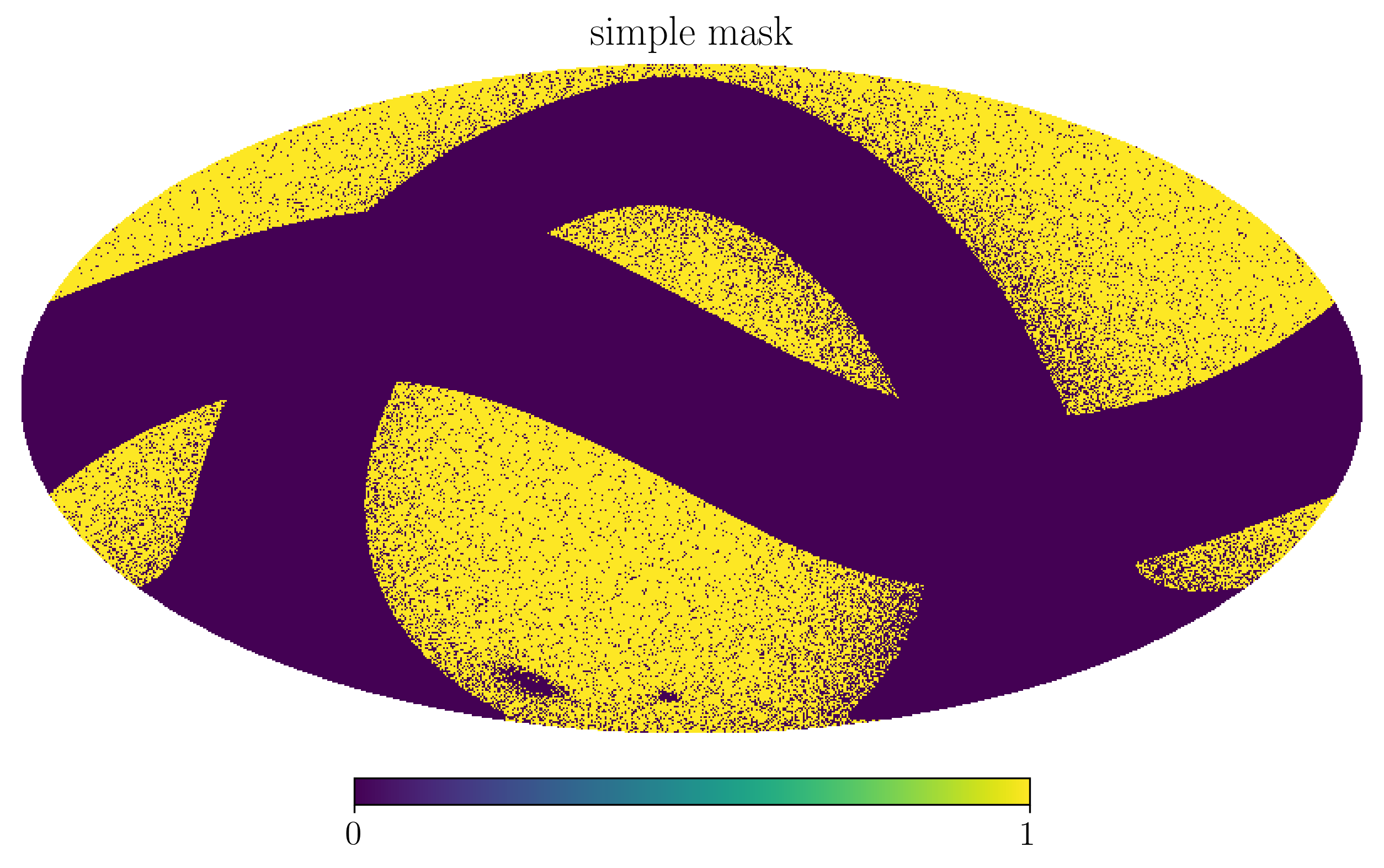}
		\caption{An illustration of the footprint for CSST cosmic shear forecast. After the masking of the galactic equator, the ecliptic equator, bright galaxies, and bright stars, the final footprint is reduced from $\sim17572\deg^2$ to $\sim14877\deg^2$. The area of this footprint will be used in the Gaussian covariance estimation (Eq.\,\ref{eq cov G}), and its geometry will be used in the super-sample covariance estimation (Eq.\,\ref{eq cov SSC}).} \label{fig mask}
	\end{figure}
	
	We also estimate how the sky coverage will change considering simple masking of the bright galaxies and bright stars. We use a \textsc{Healpix} map \citep{Healpy_Gorski2005,Healpy_Zonca2019} with $N_{\rm side}=4096$ ($\sim0.74$ arcmin$^2$ per pixel) and remove the regions within $\pm19.2\degr$ of the galactic latitude and the ecliptic latitude. The remaining region is $\sim 17572\deg^2$, which approximates the CSST target sky. We further remove bright sources, which are likely to be low-z objects with low lensing efficiency and can contaminate their nearby galaxies. We remove pixels that contain any galaxy with a magnitude brighter than 18.5 (in B band from \citealt{deVaucouleurs1991}), and any star with a magnitude brighter than 18 from GAIA DR3 \citep{Gaia2016,Gaia2022}. The resulting footprint is shown in Fig.\,\ref{fig mask}, with the sky coverage reduced to $\sim14877\deg^2$. The pixel size is significantly larger than the common size for bright object removal \citep{Coupon2018}, therefore the resulting footprint is a conservative estimation.
	
	\section{Results} \label{sec results}
	
	\begin{figure*}
		\includegraphics[width=1.5\columnwidth]{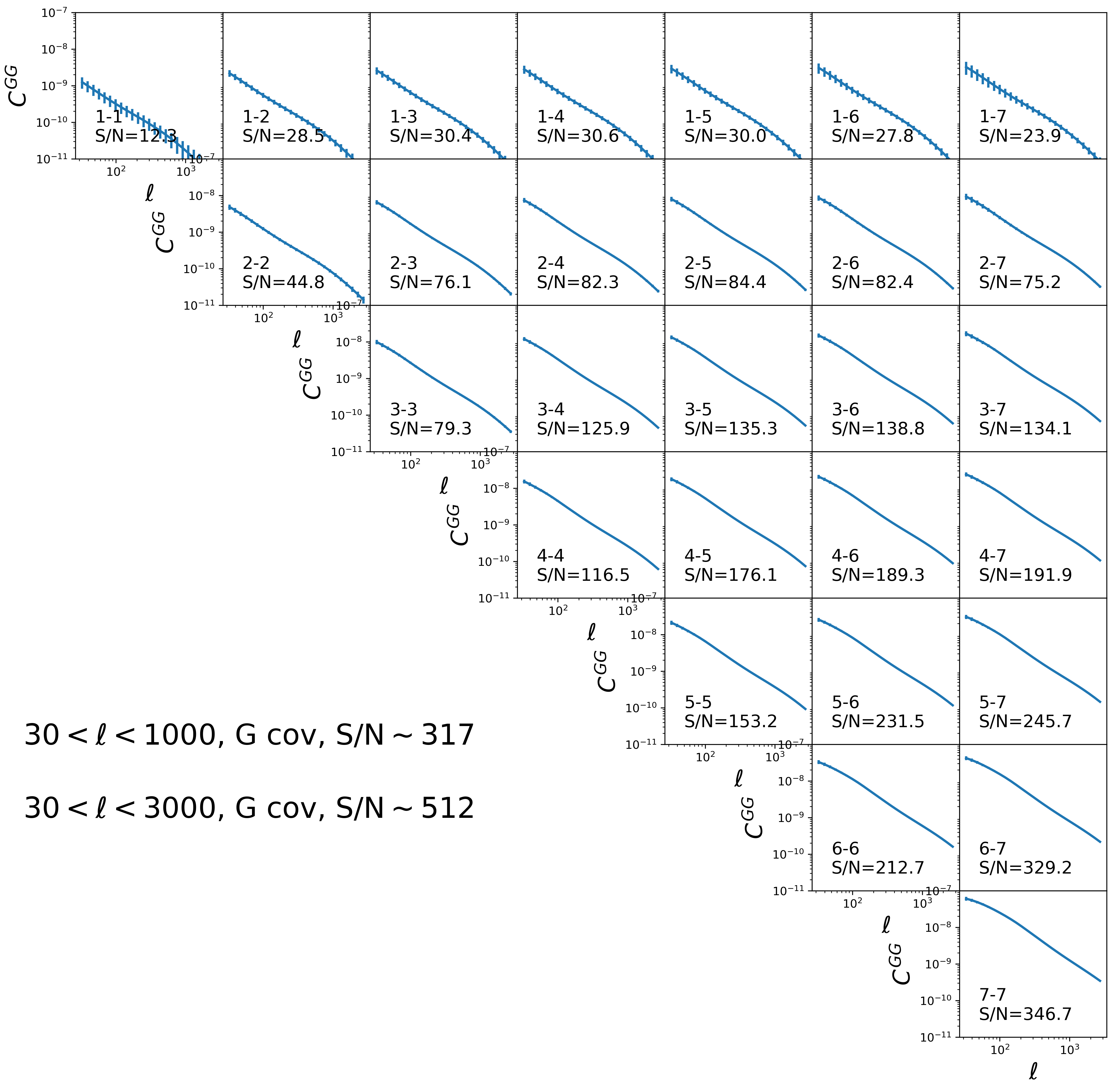}
		\caption{The tomographic shear-shear angular power spectra. In each subplot we present the $C^{\rm GG}_{ij}(\ell)$ for the $i$-th and $j$-th bins. The Gaussian errorbars are presented following Eq.\,\eqref{eq cov G} and is very small for high-z high S/N bin pairs. The S/N is evaluated for each $i-j$ pair, while the total S/N is presented with Gaussian covariance. A comparison between Gaussian covariance and non-Gaussian covariance will be shown later in Fig.\,\ref{fig cov}. } \label{fig Cell}
	\end{figure*}
	
	\begin{figure*}
		\includegraphics[width=2.0\columnwidth]{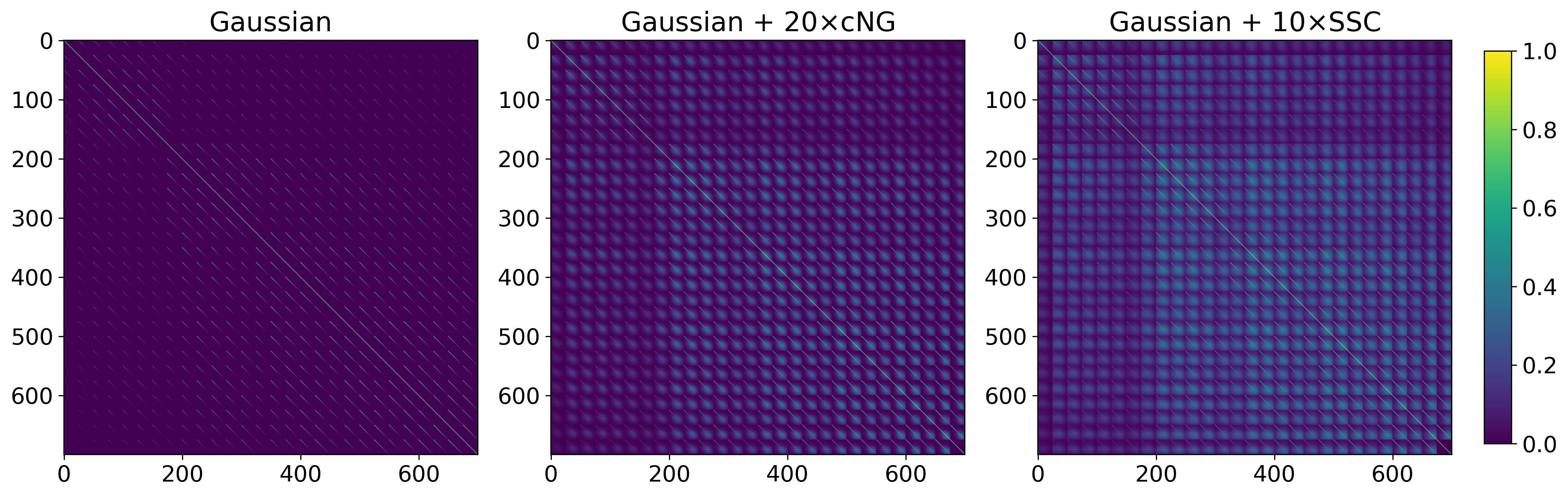}
		\caption{The impact from super-sample covariance (SSC). The left panel shows the normalized Gaussian covariance (or the correlation coefficient) for the data vector shown in Fig.\,\ref{fig Cell}, with a total length of $\sim700$ for the fiducial cut. The right panel shows the comparison when adding the contribution from the SSC. It is clear the Gaussian+SSC case still has a major diagonal feature and many small off-diagonal features that are similar to the Gaussian covariance. However, the SSC additionally introduced strong off-diagonal features for $\ell_1\ne\ell_2$ according to Eq.\,\eqref{eq cov SSC}. This suggests the SSC does not significantly enlarge the errorbars in Fig.\,\ref{fig Cell}, but the introduced off-diagonal covariance can largely reduce the total S/N. We note the contribution from connected non-Gaussian (cNG) covariance is not observable if added to the above figures. For the fiducial cut $30<\ell<3000$, consider a total covariance of Gaussian+cNG+SSC, the total S/N will be reduced from $\sim512$ to $\sim460$ due to the strong off-diagonal covariance. For the conservative cut $30<\ell<1000$, the total S/N will reduce from $\sim317$ to $\sim260$.} \label{fig cov}
	\end{figure*}
	
	\subsection{Statistics} \label{sec results statistics}
	
	\begin{figure*}
		\includegraphics[width=1.5\columnwidth]{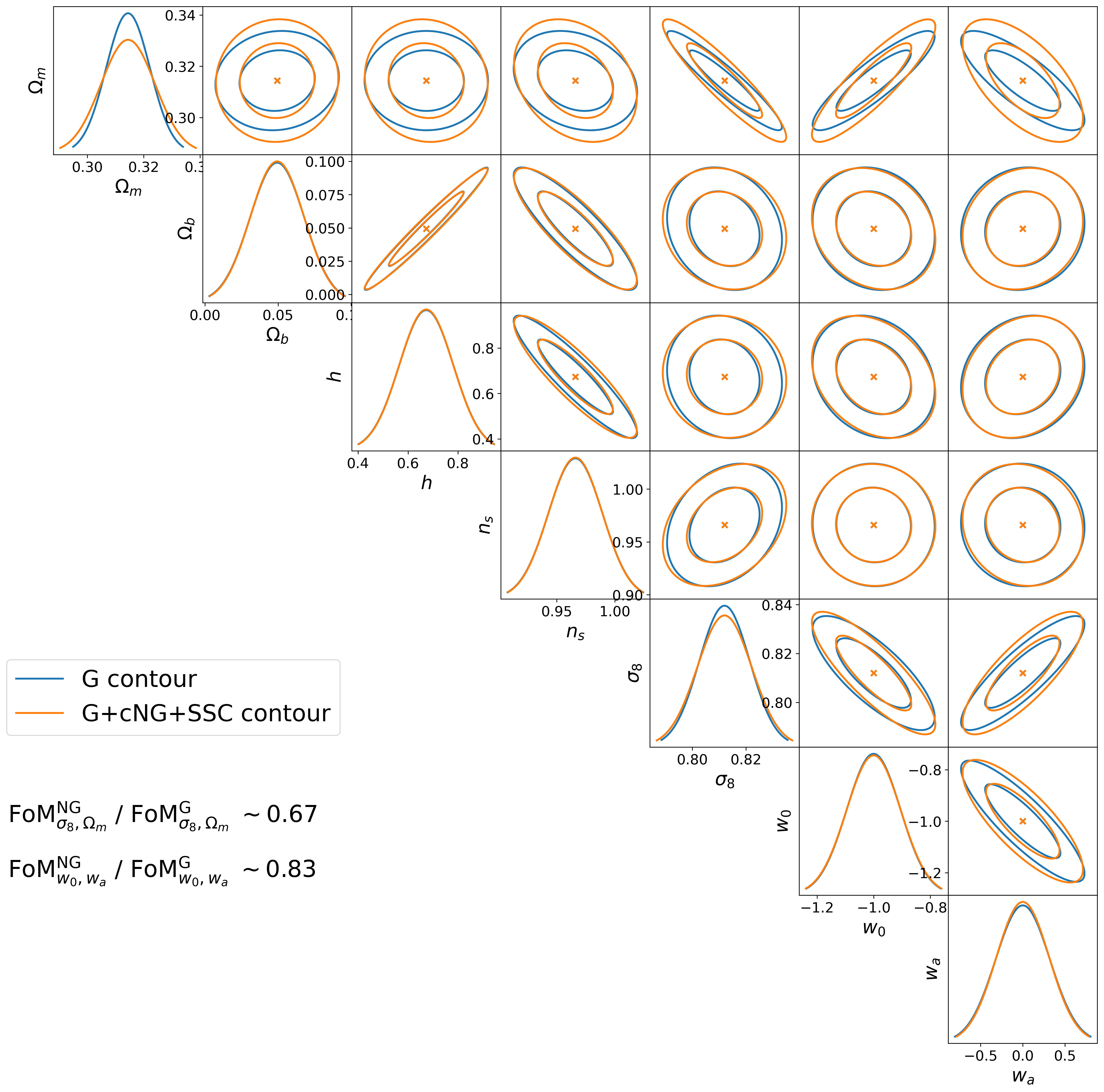}
		\caption{The forecasted cosmological constraints. The blue contours show the $1\sigma$ and $2\sigma$ confidence contours with Gaussian covariance only. When considering the full covariance, the strong off-diagonal terms in Fig.\,\ref{fig cov} will lead to a significant loss of constraining power, thus weaker constraints, shown in orange. The constraining power with the full covariance, estimated using figure-of-merit(FoM), is $\sim2/3$ in the $\sigma_8-\Omega_m$ plane and $\sim1/6$ in the $w_0-w_a$ plane comparing with the conventional Gaussian-covariance-only case. The numbers are $\sim0.70$ and $\sim0.86$ for $30<\ell<1000$.} \label{fig contour}
	\end{figure*}
	
	We first study the statistical constraining power from cosmic shear. Two different scale cuts are applied: a fiducial scale cut with $30<\ell<3000$, and a conservative scale cut with $30<\ell<1000$ in case the small scale physics are not modeled correctly. We consider logarithmic angular binning with bin width $\Delta\ell=0.2\ell$, following other stage IV surveys \citep{Yao2017}.
	
	Based on the survey properties in Sec.\,\ref{sec survey properties}, the tomographic cosmic shear angular power spectra can be calculated, shown in Fig.\,\ref{fig Cell}. Under the assumption of Gaussian covariance, we find a significantly high S/N of cosmic shear signal, with total S/N of $\sim512$ and $\sim317$ for the fiducial scale cut and the conservative scale cut, respectively.
	
	We then compare the impact from the non-Gaussian covariances, as introduced in Eq.\,\eqref{eq cov cNG} and \eqref{eq cov SSC}. We find that the connected non-Gaussian covariance has a much smaller contribution to the total covariance, compared with the Gaussian covariance and the super-sample covariance. The comparison is shown in Fig.\,\ref{fig cov}. We find that the existence of SSC does not significantly increase the errorbars in Fig.\,\ref{fig Cell}. However, it introduces a significantly strong correlation between different data points, shown as the non-diagonal terms in the Gaussian+SSC case in Fig.\,\ref{fig cov}. 
	
	The SSC is more dominant compared with the Gaussian covariance at small scales, shown in each small cube in the right panel of Fig.\,\ref{fig cov}. Each cube corresponds to the covariance between a certain $C^{\rm GG}_{ij}(\ell_1)$ v.s. $C^{\rm GG}_{mn}(\ell_2)$ pair. The diagonal feature in each cube mainly comes from the Gaussian covariance ($\ell_1=\ell_2$) and the off-diagonal features come from the SSC term. It can be seen that in many small cubes, the diagonal feature fade-away when it goes to a smaller scale (bottom-right corner). The finding of SSC being more dominant at small scales agrees with \cite{Takahashi2019}.
	
	When the total covariance including Gaussian+cNG+SSC is applied, S/N of the fiducial analysis will be reduced from $\sim512$ to $\sim460$, mainly due to the contribution from SSC. This result also agrees with \cite{Takahashi2019} that SSC is a dominate statistical error in the next stage cosmic shear studies. Nonetheless, this reduced S/N is still much stronger than the current stage III observations can achieve \citep{DESY3cosmo}.
	
	We further show the cosmological constraints considering the Gaussian covariance only as well as the full covariance in Fig.\,\ref{fig contour}. Similar to the S/N results, when considering the full covariance, the constraining power suffers from a significant loss compared with the case of using Gaussian covariance only. All the cosmological parameters, especially in the $\sigma_8$ v.s. $\Omega_m$ plane for the large-scale structure studies, and $w_0$ v.s. $w_a$ plane for the dark energy equation of state, experience enlargement in the contour due to the SSC. Some FoM values are also calculated in Fig.\,\ref{fig contour}. Overall, we conclude the impact of SSC is non-negligible for CSST cosmic shear studies.
	
	\begin{figure*}
		\includegraphics[width=1.5\columnwidth]{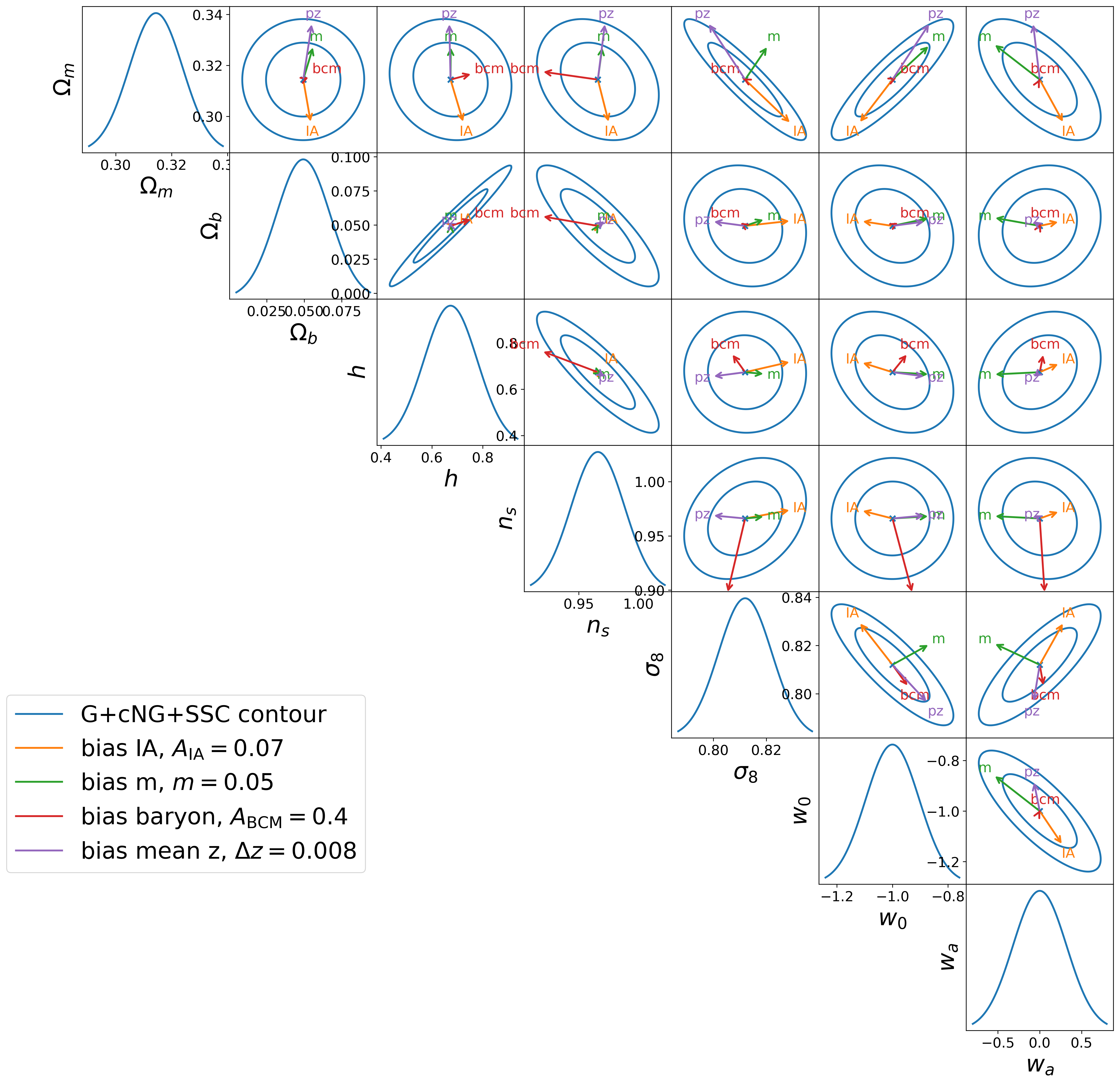}
		\caption{Systematical shifts due to different sources of residual systematics introduced in Sec.\,\ref{sec systematics}. The types and amplitudes of the residual systematics are shown in the labels, and how they can bias the cosmology, in terms of direction and amount, are shown as the arrows. The contours are the $1\sigma$ and $2\sigma$ uncertainties with the full covariance. We note this figure works as a demonstration of the systematics, and a tool to raise requirements on the systematics-control, presented in Table\,\ref{table requirement}.} \label{fig shift}
	\end{figure*}

\begin{figure*}
	\includegraphics[width=1.5\columnwidth]{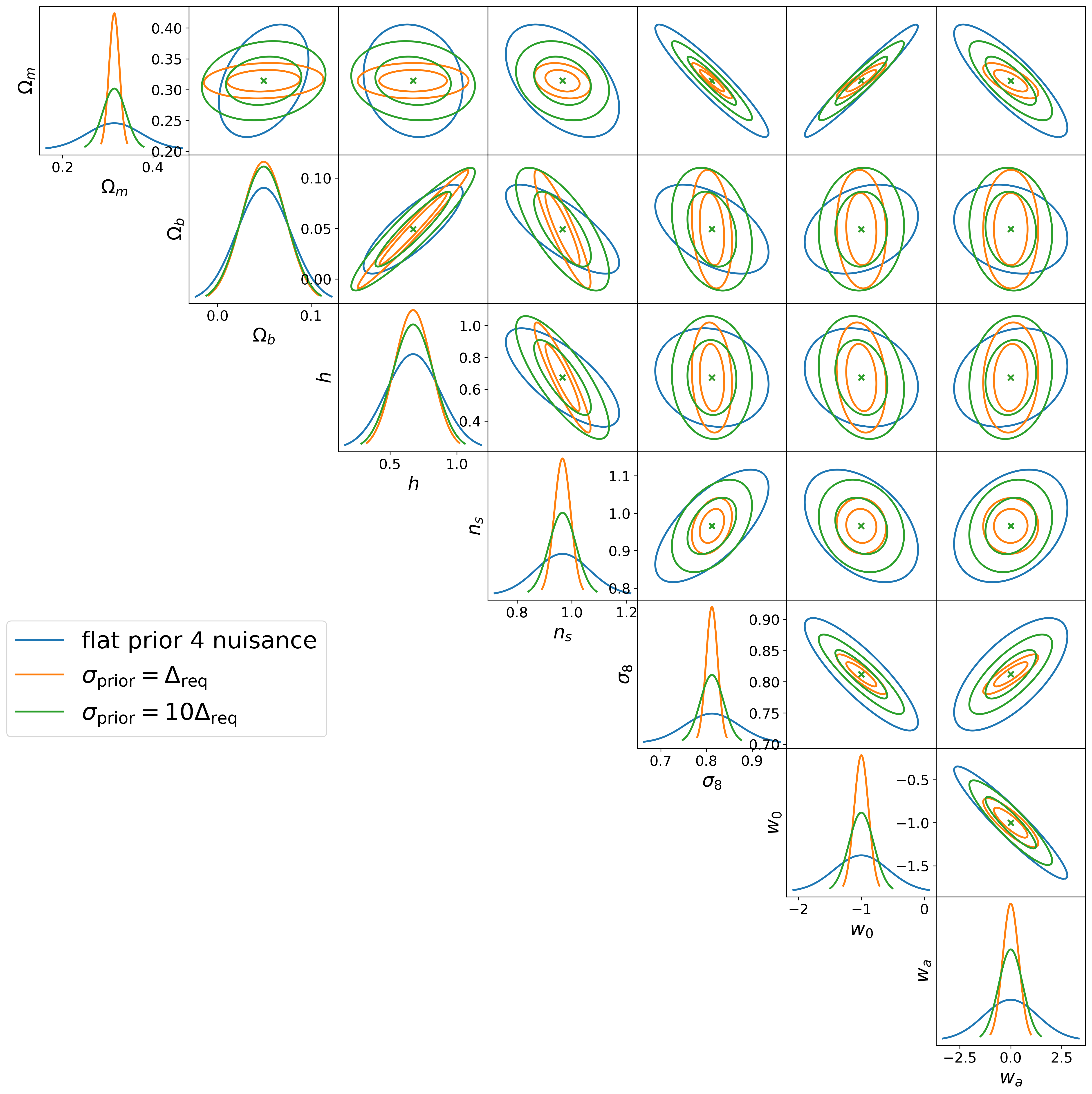}
	\caption{Constraints on the cosmological parameters after marginalization over the nuisance parameters. If we assume flat priors on the nuisance parameters, we get the blue contours (only the $68\%$ confidence level) with a huge constraining power loss in the cosmological parameters, comparing with Fig.\,\ref{fig contour} and \ref{fig shift}. If we assume Gaussian prior, with $1\sigma$ uncertainties equal to the requirements/tolerance of maximum residual bias $\Delta_{\rm req}$ at the bottom of Table\,\ref{table requirement}, we can achieve the orange contours, which are very close (maximum FoM difference $<10\%$) to the full covariance case in Fig.\,\ref{fig contour}. If the future data/simulations can not offer such a strong prior, we also tested using priors that are 10 times of the requirement, which is very close to what we can achieve today. The results are shown in green, which is about 2 times larger than the orange in each individual cosmological parameter. If this is the case, the requirements we obtained in Table\,\ref{table requirement} will also alleviate by a factor of 2.} \label{fig diff prior}
\end{figure*}
	
	\subsection{Impact of the systematics}
	
	\begin{table*}
		\centering
		\caption{Requirements for systematics-control for CSST cosmic shear. We give an example of how to read this table. If we use $30<\ell<1000$ in the angular power spectra, when the residual bias of IA reaches $A_{\rm IA}=0.05$, in $\Omega_m-\sigma_8$ plane the shift will reach the 68\% confidence contour. If we consider a bias towards this direction, it requires $A_{\rm IA}<0.01$ for the projected PDF in the bias direction to have a $>95\%$ overlap with the ideal PDF. The last row gives our requirements that correspond to previous Euclid requirements \citep{Massey2013,Euclid2011} for the fiducial CSST analysis. We refer to this requirement as $\Delta_{\rm req}\sim0.2\sigma_{\rm 2D}$. \label{table requirement}}	\begin{tabular}{ c c c c c  }
			\hline
			Case & $A_{\rm IA}$ & $A_{\rm BCM}$ & $m$ & $\Delta_z$ \\
			\hline
			$\ell<1000$, $68\%$ contour & 0.058  &  0.2  &  0.015 &  0.006 \\
			main constrain & $\Omega_{\rm m}-\sigma_8$  &  $n_s-\Omega_{\rm b}$ &  $\Omega_{\rm m}-\sigma_8$ &  $\Omega_{\rm m}-\sigma_8$, $\Omega_{\rm m}-w_a$ \\
			$\ell<1000$, $95\%$ prob & 0.012  &  0.04  &  0.003 &  0.0012 \\
			\hline
			$\ell<3000$, $68\%$ contour  ($\sigma_{\rm 2D}$) & 0.044  &  0.10  &  0.013 &  0.0042 \\
			main constrain & $\Omega_{\rm m}-\sigma_8-w_0-w_a$  &  $n_s-\Omega_{\rm b}$ &  $\Omega_{\rm m}-\sigma_8$ & $\Omega_{\rm m}-w_0$ \\
			$\ell<3000$, $95\%$ prob ($\Delta_{\rm req}$) & 0.009  &  0.02  &  0.0026 &  0.0008 \\
			\hline
		\end{tabular}
	\end{table*}
	
	We study how different systematics can bias the cosmological results, considering four different systematics we are most interested in, see Sec.\,\ref{sec systematics}. For a certain type of residual systematics, its cosmological impact is quantified by Eq.\,\eqref{eq shift}. The statistical constraints are identical to Fig.\,\ref{fig contour}, considering the full covariance with contribution from Gaussian, cNG, and SSC. {We alter the residual amplitude for each systematics so that its maximum resulting shift in all parameter spaces is at a similar level as the $2\sigma$ contour, which is easier to see.} The assumed residuals are $A_{\rm IA}=0.07$ for intrinsic alignment, $m=0.07$ for shear multiplicative bias, $A_{\rm BCM}=0.4$ for baryonic feedback, and $\Delta z=0.012$ for mean redshift.
	
	The shifts of the best-fit cosmological parameters due to the residuals of different systematic effects is shown in Fig.\,\ref{fig shift}. It is clear that different residual $\Delta C(\ell)$ can bias the cosmology towards different directions with different amounts. We then raise requirements for the systematics-control based on those biases. We define two different tolerances: a visual tolerance and a probability tolerance.\\ 
	(1) The visual tolerance, or the 68\% contour tolerance, requires all the shifts to be within the boundary of the $1\sigma$ confidence contour for a certain type of systematics. This tolerance can be directly measured by changing the residual systematics in terms of \{$A_{\rm IA}$, $m$, $A_{\rm BCM}$, $\Delta z$\}, then observing its imprint in an updated bias shift figure, similar to Fig.\,\ref{fig shift}. The actual measurements are shown in Fig.\,\ref{fig: 1sigma bias}.\\
	{(2) The probability tolerance is 0.2 of the visual tolerance. This number comes from: the length of a systematic shift that bias the cosmology right onto the edge of the $1-\sigma$ contour in 2D-space (i.e. the visual tolerance, and we refer to this length as $\sigma_{\rm 2D}$) correspond to $\sim1.52$ times the $1-\sigma$ uncertainty in the projected 1D-space ($\sigma_{\rm 1D}$) \citep{Coe2009}. While in 1D PDF, we historically require the bias to be $<0.31\sigma_{\rm 1D}$ for a 95\% overlap with the ideal 1D-PDF \citep{Massey2013}. Therefore the overall tolearance is $0.31\sigma_{\rm 1D}\frac{\sigma_{\rm 2D}}{1.52\sigma_{\rm 1D}}\sim0.2\sigma_{\rm 2D}$, which we also refer to as the 95\% probability tolerance. This definition is close to what has been used for Euclid \citep{Euclid2011,Massey2013}, but we emphasize more on bias-control for all the cosmological parameters rather than the dark energy parameters only.}
	
	We present the requirements for systematics-control for the CSST cosmic shear studies in Table\,\ref{table requirement}. For different angular scale cuts, we show the requirements on the residual systematics in terms of the visual tolerance (68\% contour) the probability tolerance (95\% prob), and in which parameter space the tolerance is triggered (contaminated the most). The one with our fiducial cuts $30<\ell<3000$ requires a 95\% probability tolerance presented at the bottom of the table, which we will refer as the requirement for CSST systematics-control $\Delta_{\rm req}$. Generally, the allowed residual systematics are at $10^{-2}$ to $10^{-3}$ level of the full contamination, which is a very strong requirement for future systematics-mitigation works.
	
	We notice the requirement on $m$ changes very little for the two different scale cuts. This is because this limitation mainly comes from the $\Omega_m-\sigma_8$ plane, and when adding the information in $1000<\ell<3000$, the contour does not change much in the $m$-bias direction in Fig.\,\ref{fig shift}. The main improvement comes from the direction that is perpendicular to the $m$-direction. Also, we note that our probability tolerance of $|m|<0.0026$ is very close to the Euclid requirement of $|m|<0.002$ \citep{Euclid2011}, considering that we introduced the non-Gaussian covariances, which magnifies the $\Omega_m-\sigma_8$ plane contour size by a factor of $\sim1.5$.
	
	We also notice that when adding the small-scale information, IA also starts to have more impact in the dark energy equation of state, becoming equivalent to the $\Omega_m-\sigma_8$ plane, see in Table\,\ref{table requirement}. This also emphasizes the importance of considering IA at small-scales and its possible deviation from the assumed tidal alignment model \citep{Blazek2015,Blazek2017,DESY3model,Shi2021,Kurita2020,Fortuna2020,Zjupa2020}.
	
	{We see that the baryonic feedback can bias \{$h$, $n_s$\} more than the other cosmological parameters. This is because its effects mainly changes the small-scale of the matter power spectrum as in Eq.\,\eqref{eq bcm Pk}. And this effect can be absorbed by the parameters that control the overall slope of the power spectrum, namely $h$ and $n_s$. This is also why by expanding the angular scale from $\ell<1000$ to $\ell<3000$, the requirement on $A_{\rm BCM}$ becomes more strict comparing with the other nuisance parameters.}
	
	\subsection{Marginalization over nuisance parameters} \label{sec marginalization}
	
	We further study the impact from marginalization over nuisance parameters, as they can steal some constraining power both due to increasing the parameter-space and due to their degeneracy with the cosmological parameters. We test 3 cases for this purpose: (1) flat priors for all the nuisance parameters; (2) if our understanding for the systematics are accurate so that the requirement of probability tolerance $\Delta_{\rm req}=0.2\sigma_{\rm 2D}$ can be achieved with compariable statistical error, so that $\sigma_{\rm prior}=\Delta_{\rm req}$; (3) if our a priori knowledge can not reach the requirement in (2), but only $\sigma_{\rm prior}=10\Delta_{\rm req}$.
	
	We note that the above case (3) uses compariable priors to what we can achieve for stage III data: \\
	($A_{\rm IA}$): For IA, one can constrain the NLA model up to $\sim10\%$ precision with simulations \citep{Hoffmann2022,Shi2021} or through self-calibration with extra observable \citep{Yao2020,Yao2023KiDS}.\\
	($A_{\rm BCM}$):  For baryonic feedback, one can constrain the signal with $\sim10\%$ level precision using the small-scale data from cosmic shear \citep{Chen2023} or using the integrated tSZ \citep{Pandey2023}, while the significance could potentially be stronger with the next stage observations \citep{ChenZY2023}.\\
	($m$ and $\Delta_z$): the current priors for the multiplicativa bias obtained from image simulations for stage III surveys are around $1\%$ level \citep{Asgari2021,DESY3cosmo,Li2023}. Most priors for redshift bias are also around $1\%$ level accuracy, based on different redshift inferences.
	
	The impact of the priors are shown in Fig.\,\ref{fig diff prior}. For case (1) with no assumed a priori knowledge, the nuisance parameters will take a large amount of the constraining power, leading to a significantly enlarged contour in blue. If the priors can achieve the level of the requirement/tolerance in Table\,\ref{table requirement}, which is our case (2), the strong degeneracy between the cosmological parameters and the nuisance parameters can be efficiently broken, leading to the orange contours which is very close to the full covariance case in Fig.\,\ref{fig contour} and \ref{fig shift}. If we consider the very pessimistic case (3) that our prior understanding for the systematics for stage IV data will not improve comparing with stage III, we end up with the green controus, which is $\sim2$ times broader than the orange case (2) in the posteriors that we are interested in ($\Omega_m$, $\sigma_8$, $w_0$, $w_a$). In this case, our requirements for the residual systematics in Table\,\ref{table requirement} can be relaxed by a factor of $\sim2$. This also states the importance in the systematics studies ---- we need to demonstrate the mitigation methods are no only accurate, but also with high significance, so that the loss in the cosmological constraints can be well controlled.
	
	\section{Conclusions}
	\label{sec conclusions}
	
	In this work, we build a realistic set-up for the future CSST cosmic shear observation, and address the problem of how non-Gaussian covariances and residual systematics can change our cosmological analysis. We consider connected non-Gaussian covariance and super-sample covariance in terms of statistics, and residual systematics from intrinsic alignment, baryonic feedback, and measurements of shear multiplicative bias and mean redshift bias from $n(z)$ reconstruction. We obtained very strong requirements on the residuals, from $10^{-2}$ to $10^{-3}$ of the assumed contamination parameters, see in Table\,\ref{table requirement}. 
	
	In terms of statistical errors, we demonstrated in Sec.\,\ref{sec results statistics} that the impact from connected non-Gaussian (cNG) covariance is small, while super-sample covariance (SSC) has an important effect that can enlarge the 2D confidence contours of some key cosmological parameters, as seen in Fig.\,\ref{fig contour}. We, therefore, emphasize the importance of taking it into consideration in future data analysis. We also suggest careful investigation of the SSC with real data considering $n(z)$ distribution and inhomogeneous galaxy distribution due to observational variation. The fact that SSC depends on the survey footprint also suggests that in order to maximize the scientific outcome for early-stage CSST data, the design of the survey strategy to minimize SSC is important.
	
	In terms of systematical errors, we considered how different systematics can bias the cosmological results in different ways, shown in Fig.\,\ref{fig shift}. We use the 2D parameter space which is mostly affected by a certain type of systematics to describe the tolerance level of the residual bias. This approach is different from the conventional analyses which normally focus on the bias in the dark energy equation of state \citep{Massey2013}. However, we still have comparable requirements in terms of shear multiplicative bias. The strong requirements for future CSST systematics-control are beyond the current constraints on those effects \citep{Yao2017,Yao2023KiDS,Schneider2015,Chen2023,Asgari2021,DESY3cosmo,Peng2022,Xu2023}. Therefore, we emphasize the importance of pushing new techniques, developing realistic simulations, and combining different approaches \citep{Alarcon2020} to further constrain those systematics.
	
	{In this analysis, we further investigate the effects of different assumed priors for the nuisance parameters. In Fig.\,\ref{fig diff prior}, we show that if the priors can reach the level of the requirement in Table\,\ref{table requirement}, the constraining-power-loss due to the nuisance parameters are negligible. If the priors are weaken by $\sim10$ times, which is compariable to the priors for stage III surveys, the constraints on the main cosmological parameters will be weaken by a factor of $\sim2$. A flat prior case is also shown for your reference (even though it is not practical). Therefore, we emphasize that the systematics-removal requires not only high accuracy, but also high significance. This means the high-fidelity simulations we use for systematics need to be large enough, and their combinations with observational methods (for example self-calibration) are also important.}
	
	The studies here primmarily concern the cosmic-shear two-point correlation analyses. To fully utilise the CSST weak lensing data for cosmological constraints, other statistical tools beyond the two-point correlations are necessary \citep{Euclid2023highorder,Shan2018,Martinet2021}. The impact of the systematics on these alternative statistics deserve further careful investigations \citep{Yuan2019,Zhang2022,Harnois-Deraps2021}.
	
	\section*{Acknowledgements}
	
	This work is supported by National Key R\&D Program of China No. 2022YFF0503403. 
	JY acknowledges the support from NSFC Grant No.12203084, the China Postdoctoral Science Foundation Grant No. 2021T140451, and the Shanghai Post-doctoral Excellence Program Grant No. 2021419. 
	HYS acknowledges the support from NSFC of China under grant 11973070, the Shanghai Committee of Science and Technology grant No.19ZR1466600 and Key Research Program of Frontier Sciences, CAS, Grant No. ZDBS-LY-7013. 
	PZ acknowledges the support of NSFC No. 11621303, the National Key R\&D Program of China 2020YFC22016.
	ZF acknowledges the support from NSFC No. 11933002 and U1931210.
	We acknowledge the support from the science research grants from the China Manned Space Project with NO. CMS-CSST-2021-A01, CMS-CSST-2021-A02, NO. CMS-CSST-2021-B01 and NO. CMS-CSST-2021-B04.
	
	We acknowledge the usage of the following packages pyccl\footnote{\url{https://github.com/LSSTDESC/CCL}, \citep{Chisari2019CCL}}, 
	healpy\footnote{\url{https://github.com/healpy/healpy}, \citep{Healpy_Gorski2005,Healpy_Zonca2019}}, 
	matplotlib\footnote{\url{https://github.com/matplotlib/matplotlib}, \citep{Hunter2007}},
	astropy\footnote{\url{https://github.com/astropy/astropy}, \citep{astropy}},
	scipy\footnote{\url{https://github.com/scipy/scipy}, \citep{scipy}}
	for their accurate and fast performance and all their contributed authors.
	
	This work has made use of data from the European Space Agency (ESA) mission
	{\it Gaia} (\url{https://www.cosmos.esa.int/gaia}), processed by the {\it Gaia}
	Data Processing and Analysis Consortium (DPAC,
	\url{https://www.cosmos.esa.int/web/gaia/dpac/consortium}). Funding for the DPAC
	has been provided by national institutions, in particular the institutions
	participating in the {\it Gaia} Multilateral Agreement.
	
	\section*{Data Availability}
	
	The data used to produce the figures in this work are availalbe through \url{https://doi.org/10.5281/zenodo.7813033}.

	
	
	\bibliographystyle{mnras}
	\bibliography{reference} 

	
	

	\appendix
	\section{The 68\% contour tolerances} \label{sec apdx 2D bias measurement}
	
\begin{figure*}
		\includegraphics[width=1.5\columnwidth]{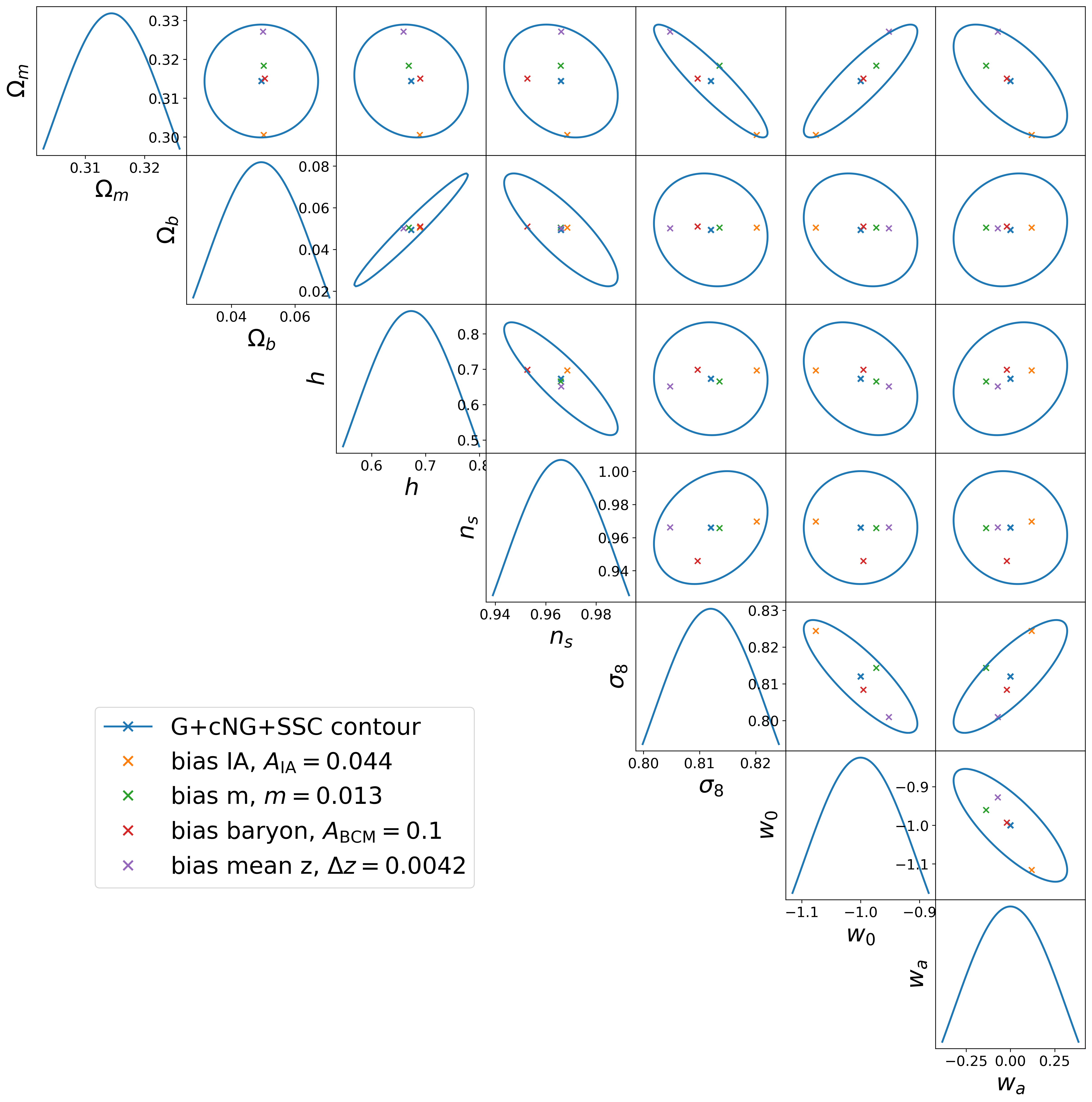}
		\caption{The 68\% contour tolerances are measured when all the shifts are within the $1\sigma$ contour, with the maximum shift right on the edge. In this figure, we use ``x'' to measure the shift, while the arrows in Fig.\,\ref{fig shift} are for presentational purposes.}
		\label{fig: 1sigma bias}
	\end{figure*}

	We show the 68\% contour tolerances for the systematics in Fig.\,\ref{fig: 1sigma bias}. The maximum shifts are right on the edge of the contours, while the values of the parameters correspond to the 68\% contour tolerances with $30<\ell<3000$ in Table\,\ref{table requirement}. We note the tips of the arrows in Fig.\,\ref{fig shift} are not very accurate, and are only for exhibition. Here we use Fig.\,\ref{fig: 1sigma bias} to show the actual values.
	
	\bsp	
	\label{lastpage}
\end{document}